\documentclass[twocolumn,superscriptaddress,aps,prc,showpacs]{revtex4}
\usepackage{amsmath,bm}
\usepackage{graphicx}

\begin{document}

\draft
%\preprint{}
\title{Scale-free download network for publications}

\author{D. D. Han} %{Ding-Ding Han}
\affiliation{Department of Electronic Engineering, East China
Normal University, Shanghai 200062, China}
\author{J. G. Liu}%Jin-Gao Liu}
\affiliation{Department of Electronic Engineering, East China
Normal University, Shanghai 200062, China}
\author{Y. G.  Ma} %{Yu-Gang Ma}
\thanks{To whom correspondence should be addressed. Email: ygma@sinr.ac.cn}
\affiliation{Shanghai Institute of Applied Physics, Chinese
Academy of Sciences, P.O. Box 800-204, Shanghai 201800, China}
\author{X. Z. Cai} %{Xiang-Zhou Cai}
\author{W. Q. Shen} %{Wen-Qing Shen}
\affiliation{Shanghai Institute of Applied Physics, Chinese
Academy of Sciences, P.O. Box 800-204, Shanghai 201800, China}

\date{\today}

\begin{abstract}
The scale-free power-law behavior of the statistics of the
download frequency of publications has been, for the first time,
reported. The data of the download frequency of publications are
taken from a well-constructed web page in the field of economic
physics (http://www.unifr.ch/econophysics/). The Zipf-law analysis
and the Tsallis entropy method were used to fit the download
frequency. It was found that the power-law exponent of
rank-ordered frequency distribution is $\gamma \sim 0.38 \pm 0.04$
which is consistent with the power-law exponent $\alpha \sim 3.37
\pm 0.45$ for the cumulated frequency distributions. Preferential
attachment model of Barabasi and Albert network has been used to
explain the download network.

\end{abstract}
¡¡¡¡
 \maketitle

Recently the complex network has become one of the hot research
fields, especially for its feature of statistical mechanics.  The
rapid growth of the internet stimulates physicists to investigate
the rules of network. In a pioneering work of Barabasi and Albert,
they found that the degree of node of Internet routes, URL
(universal resource locator) - linked networks in the WWW
(World-Wild Web) satisfies the power-law distribution
\cite{BA_Sciences,BA2}, also called as the scale-free networks.

The power-law behavior of rank distribution is believed to be
related to Zipf's law, which was found by Zipf in the early of the
last century \cite{Zipf_law}. Originally, Zipf made his remarkable
observations about some basic linguistic laws. More precisely, if
we order the words appearing in a text from the most to the less
frequent ones, we can plot the number of times of those words
appear as a function of the rank. Zipf shows that, excepting the
words with extremely low rank, an inverse power law emerges (so
called Zipf's law). That is, the frequency $x$,
\begin{equation}
x \simeq Rank^{-\gamma},
\end{equation}
where the $\gamma$ is an Zipf law exponent. Zipf's law or scale
free networks is different from the predictions of pure random
networks introduced by Erdos and Renyi \cite{Renyi}. For the
former,  Barabasi and Albert proposed a preferential attachment
model (BA model) to give the scale-free law of the link of
Internet \cite{BA_model} and Tsallis explained the statistical
feature of complex network using an non-extensive entropy (known
as Tsallis' entropy \cite{Tsallis0}) approach \cite{Tsallis}. The
original BA model predicts the probability distributions $p(k)
\simeq k^{-\alpha}$, where $k$ is the degree of network node and
$\alpha$ = 3 (the corresponding rank-ordered law yielding $\gamma$
= 1/($\alpha$ -1) = 1/2 \cite{Redner}). Extended and modified
models based on the BA model have been developed in order to
obtain $\alpha = 2-4$ more precisely to fit realistic systems
\cite{Dorogovtsev}. Recently complex networks and/or Zipf law has
been explored in very broad fields of science, which include
physics, electronics, compute sciences, geology, sociology,
economics, linguistics, biology and many others. For instance, the
complex networks have been observed for WWW and Internet, movie
actor collaboration network, science collaboration graph, cellular
networks, ecological networks, phone call networks, citation
networks, networks in linguistics, power and neural networks,
protein folding and interaction network, earthquake network, firms
growth and bankruptcy and gene expression (for a review, please
see \cite{BA_model,Dorogovtsev}), even for the fragment
hierarchical distribution in the nuclear dissociation \cite{Ma}
and the hadronic production process \cite{Wilk} etc.

The scale-free networks related to scientific publications have
been also explored, it was shown that the citation network of
scientific references \cite{Redner,Tsallis} and the collaboration
graph of the co-authorship of publications \cite{Co-author}
satisfies the power law distribution for rank-order distribution.
Redner exhibited and discussed the distributions of citations
related to two quite large data sets, namely (i) 6 716 198
citations of  783 339 papers, published in 1981 and cited between
1981 and June 1997, that have been catalogued by the Institute for
Scientific Information (ISI), and (ii) 351 872 citations, as of
June 1997, of 24 296 papers cited at least once and which were
published in Physical Review D (PRD) in volumes 11 through 50
(1975-1994). In his study, Redner addressed the citations of
publications, in variance with Laherrere and Sornette
\cite{sornette}, who addressed, in a similar study, the citations
of authors. If we denote by $x$ the number of citations and by
$N(x)$ the number of papers that are cited $x$ times. The main
results of the study were that, for relatively large values of
$x$, $N(x) \propto 1/x^{\alpha}$ with $\alpha \sim 3$, whereas,
for relatively small values of $x$, the data were reasonably well
fitted with a stretched exponential, i.e., $N(x) \propto \exp
[-(x/x_0)^{\beta}]$, $\beta$ and  $x_0$ being the fitting
parameters ($\beta \simeq 0.44$ and  $0.39$ for the ISI and the
PRD data respectively).

To help expose these differences in the citation distribution,
Redner constructed the Zipf plot \cite{Zipf_law}, in which the
number of citations of the $k$th most-ranked paper out of an
ensemble of $M$ papers is plotted versus rank $k$. By this
definition, the Zipf plot is closely related to the cumulative
large-x tail of the citation distribution and hence it is well
suited for determining the large-x tail of the citation
distribution. The integral nature of the Zipf plot also smooths
the fluctuations in the high-citation tail and thus facilitates
quantitative analysis. For the mentioned data set above, he found
that the Zipf law exponent $\gamma$ (see Eq. (1)) close to 1/2,
which is consistent with power law exponent $\alpha$ = 3 ($\alpha
= 1 + 1/\gamma$) for the distribution of citations.

In this work, we report that the rank-ordered download frequency
of the papers  in a web page can also be described by the Zipf
law. The data set we are using here comes from a well constructed
web page \cite{Zhang_web} in the field of economical physics
(so-called econophysics)  by Y. C. Zhang since 1998. The scale
free download network is explored and the quantitative information
about this complex network has been extracted by the Zipf law and
Tsallis' non-extensive entropy. The preferential attachment
network model of Barabasi and Albert is used to explain the
mechanism of network.

\begin{figure}
\vspace{-1.2truein}
\includegraphics[scale=0.4]{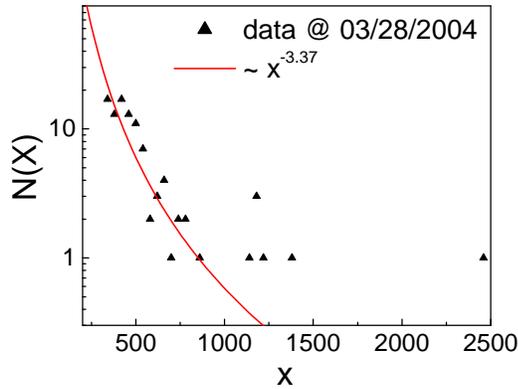}
\vspace{-1.2truein} \caption{\footnotesize The distribution of the
100 top download papers till 03/28/2004 (
http://www.unifr.ch/econophysics). $x$ represents the download
frequency and N(x) the numbers of the papers which has been
downloaded for $x$ times. } \label{fig_nx}
\end{figure}

In terms of the frequency of the download of the paper, the rank
can be defined from the most downloaded paper (rank = 1) to the
less downloaded paper. The distribution of the downloaded
frequency is shown in Fig.~\ref{fig_nx}. Roughly speaking, the
download distribution can be fitted by the power law distribution:
$N(x) \sim x^{-\alpha}$. The last few points have a large
fluctuation beyond the good fit below $x \sim 1000$. The extracted
exponent $\alpha \sim 3.37 \pm 0.45$.

\begin{figure}
\vspace{-0.2truein}
\includegraphics[scale=0.4]{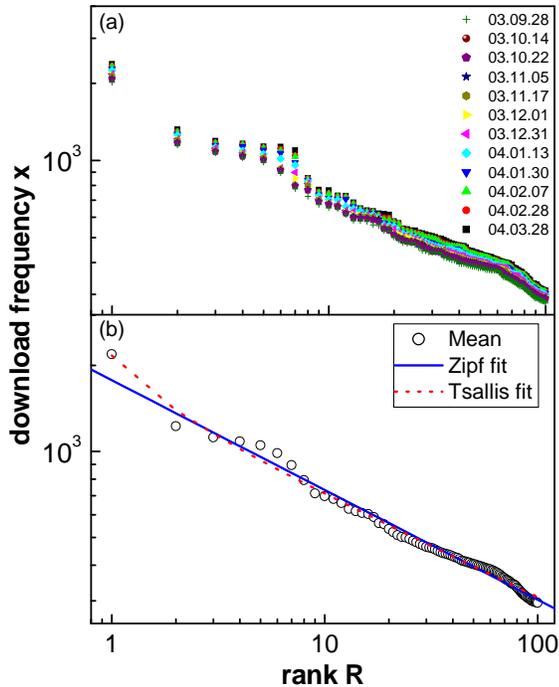}
\vspace{-0.2truein} \caption{\footnotesize The rank-ordered
(Zipf-type) plot for the download frequency  of
http://www.unifr.ch/econophysics web page. The symbols are
illustrated in figure. See text for details. } \label{fig_zipf}
\end{figure}

Since we have values of the rank-ordered frequency, we can make
the Zipf plot for  the download distribution.
Fig.~\ref{fig_zipf}(a) shows 12 Zipf plots for 12 selected dates
which are represented by the different symbols which are formatted
by Year.Month.Date. The time of these plots span from 28 Sept 2003
to 28 March  2004. In order to minimize the fluctuation of these
plots due to the statistics and network growth, we average 12 data
set points and make the Zipf plot in Fig.~\ref{fig_zipf}(b). Zipf
power law (Eq.(1)) has been used to fit the Fig. 2(b) and the
extracted exponent $\gamma \sim 0.38 \pm 0.04$. This value leads
to $\alpha = 1 + 1/\gamma = 3.63 \pm 0.28$ which is in a
reasonable agreement with $\alpha = 3.37 \pm 0.45$ from the
download distribution of Fig.~\ref{fig_nx}. It is in the range of
2 - 4 for various realistic networks \cite{Dorogovtsev}.

On the other hand, Tsallis proposed a reasonable explanation
\cite{Tsallis} and well fitted real data sets by the non-extensive
entropy theory \cite{Tsallis0}. In the non-extensive entropy
theory, the probability distribution function is given the
expectation being constant, as follows:
\begin{equation}
p(x_k) \sim \frac{1}{[1+(q-1)\lambda(x_k - \langle
 x_k\rangle)]^{\frac{q}{q-1}}},
\end{equation}
where $\langle x_k\rangle$ denotes the mathematical expectation of
$x_k$; $\lambda$ is the factor similar to Lagrange multipliers,
and $q$ is the characteristic parameter related to the exponent.
When $q$ approaches to 1, Tsallis entropy becomes the
Boltzmann-Gibbs entropy and $p(x_k)$ approaches to an exponential
distribution function.

In the rank-ordered statistics, the rank of $x_k$, $Rank(x_k)$,
and the cumulative distribution function are equivalent to a
simple relation, it reads \cite{Redner}
\begin{equation}
Rank(x_k) \propto \int_{x_k}^\infty p(x)dx = 1 - \int_0^{x_k}
p(x)dx.
\end{equation}
From integrating $p(x)$ in Eq.(2) one can
obtain the following result:
\begin{equation}
Rank \propto \frac{1}{\lambda} [1+ (q-1)\lambda (x-\langle x
\rangle)]^{-\frac{1}{q-1}};
\end{equation}
or representing $x$ as a function of $Rank$ yields
\begin{equation}
x = x_0 + \frac{b}{(Rank -Rank_{FS})^{q-1}},
\end{equation}
where $x_0$ and  $b$ are fitted parameters,
%$x_0=<x>-\frac{1}{(q-1)\lambda}$, $b \sim \frac{1}{[(q-1)\lambda^q]}$,
and parameter $Rank_{FS}$ is introduced here to take the finite
size effect into account.

By using Eq.(4) we fitted the data of average download frequency
(Fig.2(b)) and extract the parameter $q$ and $Rank_{FS}$. The
dotted line represents this fit with the parameter $q = 1.351 \pm
0.006$ and $Rank_{FS}$ = 0.60. From $q$ we deduce $\frac{q}{q-1} =
3.846 \pm 0.068$. This exponent is very close to the exponent of
$\alpha = 3.37 \pm 0.45$ in Fig.~\ref{fig_nx} as well as $\alpha =
3.63 \pm 0.28$ deduced from the $\gamma$ value of the Zipf law fit
to the rank-ordered distribution.

The scale-free characteristic of statistics of the download
frequency could be interpreted by the BA model \cite{BA_model}. In
the linear BA model, the growth of the network can be constructed
by two steps. Firstly, the earlier growth process: starting with a
small number of vertices (download papers) which visitors are
interested in them from a huge references of econophysics web
page, at every time step some visitors add some new vertices and a
rank web page of the vertices was initially constructed. Secondly,
the preferential attachment process: each visitor of econophysics
web page can freely access the rank web page of the papers. The
higher the rank of the downloaded papers, the more probability a
visitor would like to download, and the more frequency this leads
to in statistics. In this mechanism of BA model, it is natural
that such a kind of preferential attachment process will result in
a power-law or Zipf's law distribution of the downloaded
frequency.

In conclusion, the scale-free power-law behavior has been,
 for the first time, observed in the download frequency
distribution of the papers in an econophysics web page. From the
download frequency distribution, it can be described by the
power-law with the exponent $\alpha = 3.37 \pm 0.45$ which is
consistent to the description of Zipf law for the rank-ordered
download frequency with a scale-free power law exponent $\gamma
\sim 0.38 \pm 0.04$. This Zipf law parameter is not far from the
exponent from the rank-ordered citation distribution
\cite{Redner}. It may indicate of a similar mechanism for both
networks, which can be explained by the preferential attachment
process of BA model. On the other hand, the download frequency is
also considered in the framework of the non-extensive Tsallis'
entropy theory, which gives us the non-extensive Tsallis' entropy
index $q = 1.351 \pm 0.006$ and leads to $\frac{q}{q-1} = 3.846
\pm 0.068$, which  is also in well agreement with the above
$\alpha$ parameter.

%\section{Summary}

{}

\end{document}